\documentstyle[aps,twocolumn]{revtex}

\newcommand{\sa}[1]{\scriptscriptstyle{#1}}

\begin{document}

\title{The Quantum State of an Ideal Propagating Laser Field}

\author{S.J. van Enk and Christopher A. Fuchs}
\address{Bell Labs, Lucent Technologies, 600-700 Mountain Ave,
Murray Hill, NJ 07974, U.S.A.}
\date{21 November, 2001}

\maketitle

\begin{abstract}We give a quantum information-theoretic description of an
ideal propagating CW laser field and reinterpret typical
quantum-optical experiments in light of this. In particular we show
that contrary to recent claims [T. Rudolph and B. C. Sanders, Phys.\
Rev.\ Lett.\ {\bf 87}, 077903 (2001)], a conventional laser can be
used for quantum teleportation with continuous variables and for
generating continuous-variable entanglement. Optical coherence is
not required, but phase coherence is. We also show that coherent
states play a priveleged role
in the description of laser light.\end{abstract}
\medskip
A laser produces a stable, unidirectional, more or less
monochromatic, possibly very intense light beam with well-defined
coherence and polarization characteristics. These properties make a
laser a wonderful tool for optics experiments, but they are all
classical properties in the sense that they can be understood
perfectly well using Maxwell's equations. When is the quantum state
of a laser field important? As one might guess, quantum information
protocols provide examples. For instance, a recent paper by Rudolph
and Sanders~\cite{rudolph} discusses an instructive case
where---depending upon what the quantum state of a laser field is
taken to be---a laser apparently may or may not be used to
demonstrate quantum teleportation, and even may or may not be used
to generate entangled quantum states.  Their conclusion, however, is
based on an application of the standard description of a laser field
{\it inside the laser cavity}.  We show here that this is
insufficient to properly interpret various quantum information
protocols involving lasers.  As such, this provides an opportunity
to deepen our understanding of what gives quantum
information processing its power.

According to textbook laser theory---see for example \cite[ch.\
17]{lamb} and \cite[ch.\ 12]{walls}---the quantum state of the field
inside a laser cavity in a steady state is well approximated by a
mixed state diagonal in the photon-number basis. The expectation
value of the electric field in such a state vanishes. On the other
hand, many, if not all, standard optics experiments seem to be
consistent with the assumption that the laser field is in a coherent
state. The expectation value of the electric field in a coherent
state is nonzero and has a well-defined phase and amplitude. It
corresponds to a classical monochromatic light field, a solution of
the classical Maxwell equations. M\o lmer addressed the apparent
contradiction  between the two different descriptions of a laser
field in \cite{molmer}. There, he conjectured that no standard
optics experiment has yet proved the existence of a nonzero
expectation value of the electric field, and we agree with that. For
instance, he shows that a standard measurement of the phase between
two independent light beams emanating from cavities initially in
number states leads to measurement records indistinguishable from
those expected of coherent states.

The following identity is crucial for at least partly understanding
the connection between coherent-state descriptions and mixed-state
descriptions:
\begin{equation}\label{identity}
e^{-|\alpha|^2}\sum_n \frac{|\alpha|^{2n}}{n!}|n\rangle\langle n|
=\int \frac{{\rm d}\varphi}{2\pi}
|\alpha e^{i\varphi}\rangle\langle\alpha e^{i\varphi}|.
\end{equation}
The left-hand side is a mixed state diagonal in the photon-number
basis with Poissonian photon-number statistics. The right-hand side
is a mixture of coherent states with amplitude $|\alpha|$ and
arbitrary phase. We use the short-hand $\rho_{|\alpha|}$ for
this state. An experiment whose outcome does not depend on
the absolute phase $\varphi$ cannot distinguish between a pure-state
$|\alpha\rangle$ and a mixed-state $\rho_{|\alpha|}$ description.
This observation, however, is still not sufficient to fully
understand certain complicated optical experiments, as we will show
by example.

If every standard optical experiment can be described just as well
by a mixture of coherent states as by a particular coherent state,
why should one bother to find out which description is correct? It
turns out from a quantum-information theoretic point of view it
might be very important to know if one has a pure coherent state and
not a mixed state. For example, in \cite{rudolph}, Rudolph and
Sanders claim that teleportation with continuous variables is not
possible with a mixed state, but requires a true coherent state. The
main reason for their conclusion is that a mixture of two-mode
squeezed states produced by a laser in a mixed state does not
contain any entanglement. This is an important observation. In fact,
this is a splendid example of why Eq.~(\ref{identity}) does not
completely capture the essence of experiments with laser light. Here
we reexamine the question of the quantum state of a laser field from
a quantum-information theoretic perspective. Our formulation
clarifies why the coherent state plays a privileged and unique role
in the description of propagating laser fields, and how a
conventional laser can produce entanglement, even if it cannot
actually produce a two-mode squeezed state.

We consider an idealized situation where noise---in particular phase
diffusion---and transient effects are neglected. This is sufficient
for our purpose of showing that optical coherence (that is, nonzero
off-diagonal matrix elements of the density matrix in the
number-state basis) is not required for teleportation with
continuous variables. In a separate paper \cite{vanEnkQIC} we will
consider the quantum state of a realistic laser beam as well as a
more detailed account of the idealized case.

We model the laser as a one-sided cavity driven by a constant force
(a voltage or an external field) far above threshold. As is well
known from standard laser theory \cite{lamb,walls} the steady-state
density matrix of the field inside the laser cavity is diagonal in
the number-state basis with Poissonian photon number statistics. 
For convenience we first assume that the field is in a coherent
state and calculate the quantum state of the field outside the laser
cavity. Subsequently, using the identity (\ref{identity}), we 
adapt that result to find the quantum state of a propagating laser
beam of a laser in the proper mixed state.

We employ standard input-output theory \cite{walls,collett} to
connect the quantum field inside a laser cavity to its output field.
First, we separate the field modes into two parts. A single-mode
annihilation operator $a$ describes the field with frequency
$\omega_0$ inside the cavity; continuous-mode operators $b(\omega)$
describe modes with frequency $\omega$ outside the cavity. We define
input and output operators by
\begin{eqnarray}
a_{{\rm in}}(t) &=& \frac{-1}{\sqrt{2\pi}}\int{\rm d}
\omega e^{-i\omega (t-t_0)}b_0(\omega),\nonumber\\
a_{{\rm out}}(t) &=& \frac{1}{\sqrt{2\pi}}\int{\rm d}\omega
e^{-i\omega (t-t_1)}b_1(\omega),
\end{eqnarray}
where $t_0\rightarrow -\infty$ is a time in the far past and
$t_1\rightarrow\infty$ is a time in the far future. The operators
$b_0(\omega)$ and $b_1(\omega)$ are defined to be the Heisenberg
operators $b(\omega)$ at times $t=t_0$ and $t=t_1$, respectively.
The input and output operators satisfy the proper bosonic
commutation relations for continuous-mode operators,
$[a_{{\rm in,out}}(t),a^{\dagger}_{{\rm in,out}}(t')]=\delta(t-t')$.
The relation
\begin{equation}\label{boundary}
a_{{\rm in}}(t)+a_{{\rm out}}(t)=\sqrt{\kappa}a(t),
\end{equation}
with $\kappa$ the decay rate of the cavity, may be regarded as a
boundary condition on the electric field. When the input field is
the vacuum and the field inside the cavity is a coherent state
$|\alpha e^{-i\omega_0t}e^{i\phi}\rangle$, then according to
(\ref{boundary}) the output field is an eigenstate of $a_{{\rm
out}}$ with eigenvalue $\beta\equiv \sqrt{\kappa}\alpha
e^{-i\omega_0t}e^{i\phi}$. Such a state is a continuous-mode
coherent state \cite{loudon} and can be written in the Schr\"odinger
picture as
\begin{equation}\label{coco}
|\{\beta(t)\}\rangle\equiv\exp\!\left(\int{\rm d}\omega [
\beta(\omega) b^{\dagger}(\omega) -
\beta^*(\omega)b(\omega)]\right)|{\rm vac}\rangle,
\end{equation}
with $|{\rm vac}\rangle$ the vacuum state and $\beta(\omega)$ the
Fourier transform of $\beta(t)$. A continuous-mode coherent state
can be described alternatively as an infinite tensor product of
discrete-mode coherent states \cite{loudon}.  Let $\{\Phi_i(t)\}$ be
a set of functions satisfying the orthogonality and completeness
relations,
\begin{eqnarray}\label{complete}
\int {\rm d}\tau \Phi_i(\tau) \Phi_j^*(\tau)&=&\delta_{ij},\nonumber\\
\sum_i \Phi_i(t)\Phi_i^*(t')&=&\delta(t-t').
\end{eqnarray}
We may then define annihilation operators $c_i$ (satisfying the
correct bosonic commutation relations for discrete operators)
according to $c_i=\int {\rm d}t \Phi^*_i(t) a_{{\rm out}}(t)$. An
eigenstate of $a_{{\rm out}}(t)$ with eigenvalue $\beta(t)$ is also
an eigenstate of $c_i$ with eigenvalue $\alpha_i=\int {\rm
d}t\Phi^*_i(t) \beta(t)$. We now apply this formalism to describe
laser light as a sequence of packets of light, each with the same
duration $T$. Let the functions $\{ \Psi_n(t) \}$ be defined by
\begin{eqnarray}\label{Psin}
\Psi_n(t)&=&\frac{\exp(-i\omega_0t)}{\sqrt{T}}\,\,\,
{\rm for}\,\,\, \left|t-\frac{z_0}{c}-nT\right|<\frac{T}{2},\nonumber\\
&=&0 \,\,\,{\rm otherwise}.
\end{eqnarray}
The label $z_0$ refers to an arbitrarily chosen reference position
relative to which we partitioned the light beam into equal pieces of
length $cT$. This set of functions is orthogonal and can be extended
to form a complete set satisfying (\ref{complete}). For a CW laser
described by $\beta(t)=\sqrt{\kappa}\alpha e^{-i\omega_0t}e^{i\phi}$
we see that each part $n$ of the light beam is in the same coherent
state with eigenvalue $\alpha_n=\sqrt{\kappa T} \alpha
e^{i\phi}\equiv \alpha_0$, corresponding to the modes described by
(\ref{Psin}), and $\alpha_i=0$ for all other modes.

Now assuming that the field inside the laser cavity is in fact a
mixture $\rho_{|\alpha|}$, the quantum state of a sequence of $N$
parts corresponding to the set $\{ \Psi_n\}$ is thus
\begin{equation}\label{ensemble}
\tilde{\rho}_N =\int \frac{{\rm d}\varphi}{2\pi}\big( |\alpha_0
e^{i\varphi}\rangle \langle \alpha_0 e^{i\varphi}|\big)^{\otimes
N}\;,
\end{equation}
where the integrand signifies an $N$-fold tensor product over the
separate packets.

This result\cite{beamsplitter} displays a privileged role for
coherent states in describing a propagating laser field: Although
the quantum state inside the laser is a mixed state diagonal in the
number-state basis, the quantum state of the output is not equal to
a product of mixed states $(\rho_{|\alpha_0|})^{\otimes N}$ (it
would be for a pulsed laser). Rather it can be thought of as a mixture of $N$ copies
of a coherent state, each copy with the same ``unknown'' phase. The
real question is, is this the only such description?  We would
certainly not want to commit the {\it preferred ensemble fallacy\/}
(PEF) that Rudolph and Sanders~\cite{rudolph} rightly warn of.

The answer is given by the quantum de Finetti
theorem~\cite{hudson,caves}:  Consider a source producing a sequence
of systems with the property that interchanging any two of the
systems will not change the joint probability distribution for the
outcomes of measurements on the individuals\cite{noise}. Moreover, suppose this
exchangeability property holds even when the ensemble is extended by
any number of systems. The quantum de Finetti representation theorem
specifies that the quantum state of any $N$ systems from such a
source is {\it necessarily\/} of the form
\begin{equation}\label{Finetti}
\tilde{\rho}_N=\int {\rm d}\rho P(\rho) \rho^{\otimes N}\;,
\end{equation}
where $P(\rho)$ is a probability distribution over the density
operators and ${\rm d}\rho$ is a measure on that space. Most
importantly, this representation is unique up the behavior of
$P(\rho)$ on a set of measure zero.

Now contemplate performing a set of measurements on the individual
systems emanating from our source. The probability distribution
$P(\rho)$ in (\ref{Finetti}) must be updated according to standard
Bayesian rules after the acquisition of that
information~\cite{schack}. Indeed, if the measurements are performed
on a sufficiently large subset, and the measurements form a complete
set in the space of operators, then the probability distribution
will tend to a delta function $P(\rho)\rightarrow
\delta(\rho-\rho_0)$. Comparing the state of a propagating laser
field (\ref{ensemble}) with the general form (\ref{Finetti}) we see
that a {\em complete\/} set of measurements on part of the light
emanating from the laser will reduce the quantum state of the rest
of the light to a pure state, and this pure state will necessarily
be a {\em coherent state}. This shows the unique role of coherent
states in the description of laser light.

It is true that standard optics experiments have not yet featured
such complete measurements. For instance, a complete set for the case
at hand would be a measurement of amplitude and absolute phase.
However, recent developments \cite{jones} may make it possible to
compare the phase of an optical light beam directly to the phase of
a microwave field. Using this technique the only further measurement
required for a complete measurement is a measurement of the absolute
phase of the microwave field, which is possible electronically. This
measurement would create an optical coherent state from a standard
laser source for the first time. But as we will show in the next
section, such a measurement does not need to be performed for most
applications.

Let us now describe a typical optical experiment using
(\ref{ensemble}) for a proper description of the quantum state of a
laser.
M\o lmer in
\cite{molmer} showed that the detection of a phase difference
between two (independent) light beams need not imply that there is a
well-defined phase difference before the measurement. In particular,
he showed that for light emanating from two cavities whose fields
are initially in number states (whose phase is completely random),
the standard setup to measure phase will indeed find a stable phase
difference (though the value of this phase will be random and
different from experiment to experiment). Within one experiment, it
takes just a few (about three) photon detections \cite{molmer} to
settle on a particular value of the phase difference, after which the
counting rates of the detectors remain consistent with that initial
phase difference. In other words, the standard phase measurement
acts almost like a perfect von Neumann measurement; the measurement
will produce an eigenvalue of the corresponding observable and the
state after the measurement can be described by an eigenstate of the
measured variable.
Generalizing this observation to continuously pumped CW lasers leads
to the following simple description. Initially we have two
independent laser beams $A$ and $B$ whose joint quantum state is
described by
\begin{eqnarray}\label{2}
\tilde{\rho}_{2N} &=& \int \frac{{\rm d}\varphi_A}{2\pi}\big(
|\alpha_{\sa{A}} e^{i\varphi_{\sa{A}}}\rangle \langle \alpha_{\sa{A}}
e^{i\varphi_{\sa{A}}}|\big)^{\otimes N}
\nonumber\\
&& \otimes \int \frac{{\rm d}\varphi_{\sa{B}}}{2\pi}
\big( |\alpha_{\sa{B}} e^{i\varphi_{\sa{B}}}\rangle \langle
\alpha_{\sa{B}} e^{i\varphi_{\sa{B}}}|\big)^{\otimes N}
\end{eqnarray}
if we divide each laser beam into $N$ packages of constant duration.
If the first package of each beam is used to measure a phase
difference then the state of the rest of the light beams will be
reduced to
\begin{eqnarray}\label{2'}
\tilde{\rho}_{2N-2} &=& \int \frac{{\rm d}\varphi_A}{2\pi}\big(
|\alpha_{\sa{A}} e^{i\varphi_{\sa{A}}}\rangle \langle \alpha_{\sa{A}}
e^{i\varphi_{\sa{A}}}|
\big)^{\otimes (N-1)}\nonumber\\
&& \otimes \big( |\alpha_{\sa{B}} e^{i(\phi_0+\varphi_{\sa{A}})}\rangle
\langle \alpha_{\sa{B}} e^{i(\phi_0+\varphi_{\sa{A}})}|\big)^{\otimes
(N-1)},
\end{eqnarray}
where we assumed the outcome of the phase measurement was $\phi_0$
and approximated the measurement to be sharp. The state (\ref{2'})
has the property that a subsequent measurement of the phase
difference will reproduce the value $\phi_0$: This is a kind of
``phase-locking without phase.'' Note this would certainly not be
the case if the quantum state of a laser were a product of identical
mixed states of the form $(\rho_{|\alpha|})^{\otimes N}$.

We now address the issue of teleportation with continuous variables
using a two-mode squeezed state \cite{akira}.
Such a state can be generated by splitting two squeezed states on a 50-50
beamsplitter. The resulting state of the two output ports is an
entangled state. Denote a two-mode squeezed state generated from a
coherent state with amplitude $\alpha e^{i\varphi}$ by
$|T^{\sa{AB}}_\alpha(\varphi)\rangle$, where the superscripts $A,B$
refer to two distinct modes located in different laboratories, say
Alice's and Bob's. As shown in \cite{rudolph}, the state
\begin{equation}\label{t}
\int \frac{{\rm d}\varphi}{2\pi} |T^{\sa{AB}}_\alpha(\varphi)
\rangle\langle T^{\sa{AB}}_\alpha(\varphi)|
\end{equation}
contains no entanglement between $A$ and $B$: Instead, it simply
denotes classical correlation between photon numbers for the two
modes.
Now, however, suppose that some of the remaining laser light is
supplied to Alice (as for instance for the purpose of producing a
local oscillator \cite{akira}).  The overall quantum state between
Alice and Bob will then be of the form
\begin{equation}\label{t2}
\int \frac{{\rm d}\varphi}{2\pi} |T^{\sa{AB}}_\alpha(\varphi)\rangle
\langle T^{\sa{AB}}_\alpha(\varphi)|\otimes \big( |\alpha_{\sa{A'}}
e^{i\varphi}\rangle\langle \alpha_{\sa{A'}}
e^{i\varphi}|\big)^{\otimes N},
\end{equation}
where $A'$ indicates the further modes in Alice's possession. Far
from being an unentangled state, this state has every bit as much
entanglement as if the laser were actually a pure coherent source.
It is just that the entanglement is in the form of {\it distillable
entanglement\/} \cite{ent}.
To see this, contemplate Alice doing a complete measurement on the
extra laser light in her lab.  With it, she will reduce the quantum
state of modes $A,B$ to a true two-mode squeezed state. Since these
measurements are local (all measurements are performed on Alice's
modes $A'$), it follows there must be distillable entanglement
between Alice's and Bob's modes.  Although the claim in
\cite{rudolph} that the state (\ref{t}) can be produced locally by
Alice and Bob is quite correct, the state (\ref{t2}) is entangled
and cannot be so produced.

This shows that teleportation of continuous variables is possible
even with lasers in mixed states. The actual procedure used in
\cite{akira} required, as was noted in \cite{rudolph}, both Alice
and Bob to use some of the light of the same laser that generated
the two-mode squeezed state to perform homodyne detection. The fact
that Bob shares laser light with Alice does not imply however, that
they share any quantum channel over and above their original
entanglement.  In principle all the light in Alice and Bob's
possession (both the shared two-mode squeezed state and the light
for their local oscillators) was sent to them {\em before\/} any
actual teleportation takes place.

Moreover, as pointed out in \cite{enk}, such a shared resource is
necessary for any teleportation protocol, irrespective of its
physical implementation. For teleportation with continuous
variables, Alice and Bob need to share a synchronized clock; sharing
some of the laser light is a practical way of implementing this (though 
of course laser light is more than simply a clock ).
In
contrast to \cite{rudolph}, we do not consider the presence of this
resource, which acts as a phase reference, as invalidating
teleportation. An independent party, Victor, who would
like to verify Alice and Bob's teleportation skills, could use his
own laser but has to ``phase-lock'' it with Alice's laser
\cite{frequency}. After all, Alice's claim is only that she can
teleport a quantum state of a particular mode: Victor is free to
choose the state to be teleported, but not the Hilbert space.

Finally, the teleportation procedure as a whole does not depend on
the value of the absolute phase $\varphi$. Therefore, for
teleportation to succeed, Alice does not even have to do an absolute
phase measurement to actually distill the entanglement present in
the state (\ref{t2}). Teleportation can be achieved without knowing
the imagined ``unknown'' phase $\varphi$ arising in any PEF. In
particular, Alice and Bob can teleport a quantum state handed to
them by the independent third party Victor even if he is able to
generate a pure coherent state or a pure entangled state. This is
because the phases of both input and output state are compared to
one and the same phase reference. Of course, in the actual
experiment \cite{akira} no coherent state was produced and thus no
coherent states were teleported. Instead it is the action of a
general displacement operator (which acts on a coherent state as
$D_\beta|\alpha\rangle=\exp(i{\rm
Im}(\beta\alpha^*))|\alpha+\beta\rangle$) that is teleported.

In conclusion, viewing the laser beam of a CW laser as a sequence of
$N$ quantum systems leads to the following result: The quantum state
of a laser beam is a mixture of $N$ copies of identical pure coherent
states. Such a state is very different from $N$ copies of identical
mixed states (be they mixtures of number states or of coherent
states). One consequence is that appropriate measurements performed
on part of a laser beam will reduce the quantum state of the rest of
the laser beam to a pure coherent state. Such measurements seem in
fact possible with present-day technology \cite{jones}, and thus an
optical coherent state may in fact be generated.  No sophisticated
measurement on the laser medium \cite{molmer} need be contemplated
to carry this out.

Most importantly, this description allows us to properly assess
quantum communication protocols that rely on lasers. In particular
we find that teleportation with continuous variables is possible with
conventional lasers without actually having to reduce the quantum
state of a laser to a coherent state.

We thank Terry Rudolph and Barry Sanders for sending us a
copy of their paper, and Klaus M\o lmer and the referees 
for extended comments.


\begin{thebibliography}{99}
\bibitem{noise}Such exchangebility of course would not hold for a phase-diffusing laser.
However, as already stated, consideration of the idealized case is sufficient for tackling
the issues considered in \cite{rudolph,molmer}.
\bibitem{frequency}
Actually, if there is a guarantee that both lasers have the same
frequency, no phase-locking is even necessary: simply a fixed phase
relation between the lasers is sufficient.
\end{thebibliography}
\end{document}